\documentclass[nonlinearer, 9pt, conference]{ieeeconf}      

\IEEEoverridecommandlockouts                              

\usepackage{graphics} 
\usepackage{epsfig} 
\usepackage{times} 
\usepackage{amsmath} 
\usepackage{amssymb}  
\usepackage{subfigure}
\usepackage{url}
\usepackage{bbm, dsfont}
\usepackage{mathtools}
\usepackage{multirow}
\newtheorem{theorem}{Theorem}

\newtheorem{algorithm}[theorem]{Algorithm}

\usepackage{cite}

\usepackage{color}
\usepackage{algorithm}

\newtheorem{definition}[theorem]{Definition}

{ 
\newtheorem{remark}[theorem]{Remark}

}

\newcommand{\bi}{\begin{itemize}}
\newcommand{\ei}{\end{itemize}}
\newcommand{\bd}{\begin{displaymath}}
\newcommand{\ed}{\end{displaymath}}
\newcommand{\be}{\begin{eqnarray*}}
\newcommand{\ee}{\end{eqnarray*}}

\newcommand{\K}{{\bf K}}
\usepackage[normalem]{ulem}

\title{\LARGE \bf
Data Driven Online Learning of Power System Dynamics}
\author{Subhrajit Sinha, Sai Pushpak Nandanoori and Enoch Yeung\\
\thanks{The Pacific Northwest National Laboratory (PNNL) is operated by Battelle for the U.S. Department of Energy under Contract DE-AC05-76RL01830.}
\thanks{This work was supported partially by a Defense Advanced Research Projects Agency (DARPA) Grant No. DEAC0576RL01830 and an Institute of Collaborative Biotechnologies Grant.}
\thanks{S. Sinha and S. P. Nandanoori are with PNNL, Richland, WA 99354 USA (emails: subhrajit.sinha@pnnl.gov, saipushpak.n@pnnl.gov),  and E. Yeung is with University of California Santa Barbara, CA 93106 (email: eyeung@ucsb.edu)
}
}

\begin{document}
\maketitle

\begin{abstract}
With the advancement of sensing and communication in power networks, high-frequency real-time data from a power network can be used as a resource to develop better monitoring capabilities. In this work, a systematic approach based on data-driven operator theoretic methods involving Koopman operator is proposed for the online identification of power system dynamics. In particular, a new algorithm is provided, which unlike any previously existing algorithms, updates the Koopman operator iteratively as new data points are acquired. The proposed algorithm has three advantages: a) allows for real-time monitoring of the power system dynamics b) linear power system dynamics (this linear system is usually in a higher dimensional feature space and is not same as linearization of the underlying nonlinear dynamics) and c) computationally fast and less intensive when compared to the popular Extended Dynamic Mode Decomposition (EDMD) algorithm. The efficiency of the proposed algorithm is illustrated on an IEEE 9 bus system using synthetic data from the nonlinear model and on IEEE 39 bus system using synthetic data from the linearized model. 
%
\end{abstract}

\section{Introduction}\label{section_introduction}
The modern power grid is a cyber-physical system (CPS) with generators, loads, transmission lines constituting the physical elements of CPS and the cyber layer connects the physical system to the control center where sensor measurements and actuation signals are communicated. With advancements in sensing and actuation, this CPS paradigm is increasingly viewed as a critical infrastructure for achieving reliable operation of the power grid. This can be achieved with better real-time monitoring and feedback control strategies \cite{de2010synchronized,pushpak2016control}. This work deals with identifying the real-time power system dynamics from data that can help in achieving better real-time monitoring capabilities.

In recent years, with advances in information technology and data-processing capabilities, there has been an increased interest in data-driven analysis in almost all different areas of science and engineering. In the realm of dynamical systems, operator theoretic methods for data-driven analysis of dynamical systems has gained particular importance in recent years \cite{Dellnitz_Junge,Mezic2000,froyland_extracting,Junge_Osinga,Mezic_comparison,
Dellnitztransport,mezic2005spectral,Mehta_comparsion_cdc,Vaidya_TAC,
raghunathan2014optimal,susuki2011nonlinear,mezic_koopmanism,
mezic_koopman_stability,yeung2015global, yeung2018koopman,yeung2017learning, sparse_Koopman_acc,johnson2018class,sinha_online_arxiv,sai_phase_space_arxiv}. The operator theoretic methods, involving Perron-Frobenius (P-F) and Koopman operators have the advantage that even if the underlying system is nonlinear, the evolution of these operators is linear. However, usually, these operators are linear operators in an infinite-dimensional space and thus one needs to compute finite-dimensional approximations of these operators. To this end, there are different algorithms for Koopman operator computation, with Dynamic Mode Decomposition (DMD) \cite{DMD_schmitt} and Extended Dynamic Mode Decomposition \cite{EDMD_williams} being the most popular ones.

However, all these algorithms use the entire obtained data-set for computing the finite-dimensional approximation of the Koopman operator. Hence, if a new data point is acquired and the Koopman operator needs to be updated, the Koopman operator has to be computed from scratch using the updated data-set. This makes the existing algorithms computationally intensive and thus impractical for real-time identification and monitoring of dynamical systems. 

In this paper, we propose an iterative algorithm for the computation of the Koopman operator. In particular, when a new data point comes in, the Koopman operator, which is already known at the previous time instant, is updated incrementally using the new data point. As this algorithm uses only the existing Koopman operator and just the new data point, as opposed to the previously existing algorithms it is computationally much more efficient. Hence, this is more suited for real-time identification of dynamical systems. To this end, we show how the proposed algorithm can be used to identify the power system dynamics in real-time in a computationally efficient manner.

The paper is organized as follows. In section \ref{section_Koopman}, we discuss the basics of transfer operators followed by discussion of EDMD algorithm and recursive Koopman learning in section \ref{section_online_Koopman}. The design of a robust predictor based on learned Koopman operator is presented in section \ref{section_predictor}. Simulations on two IEEE test cases are presented in section \ref{section_simulations} and the paper concludes in section \ref{section_conclusion}.


\section{Koopman Operator}\label{section_Koopman}
Consider a discrete-time system
\begin{equation}\label{system}
z_{t+1} = T(z_t)
\end{equation}
where $z_t\in\mathbb{R}^N$ and $T: M \subset \mathbb{R}^N\rightarrow M$ is assumed to be an invertible smooth diffeomorphism. Typically the system (\ref{system}) is analysed by studying the solution of the difference equation (\ref{system}). For continuous-time systems, the analysis is done by studying the solutions of the corresponding differential equation. However, associated with the dynamical system (\ref{system}) are two operators, namely Perron-Frobenius (P-F) $(\mathbb{P})$ and Koopman operators $(\mathbb{U})$, which study the evolution of functions under the mapping $T$. 

The Koopman operator associated with the dynamical system (\ref{system}) is defined as follows \cite{Lasota}:
\begin{definition} [Koopman Operator] 
Given any $g\in\cal{F}$, $\mathbb{U}:{\cal F}\to {\cal F}$ is defined by
\[[\mathbb{U} g](z)=g(T(z)),\]
where $\cal F$ is the space of function (observables) invariant under the action of the Koopman operator.
\end{definition}

\begin{figure}[htp!]
\centering
\includegraphics[scale=.25]{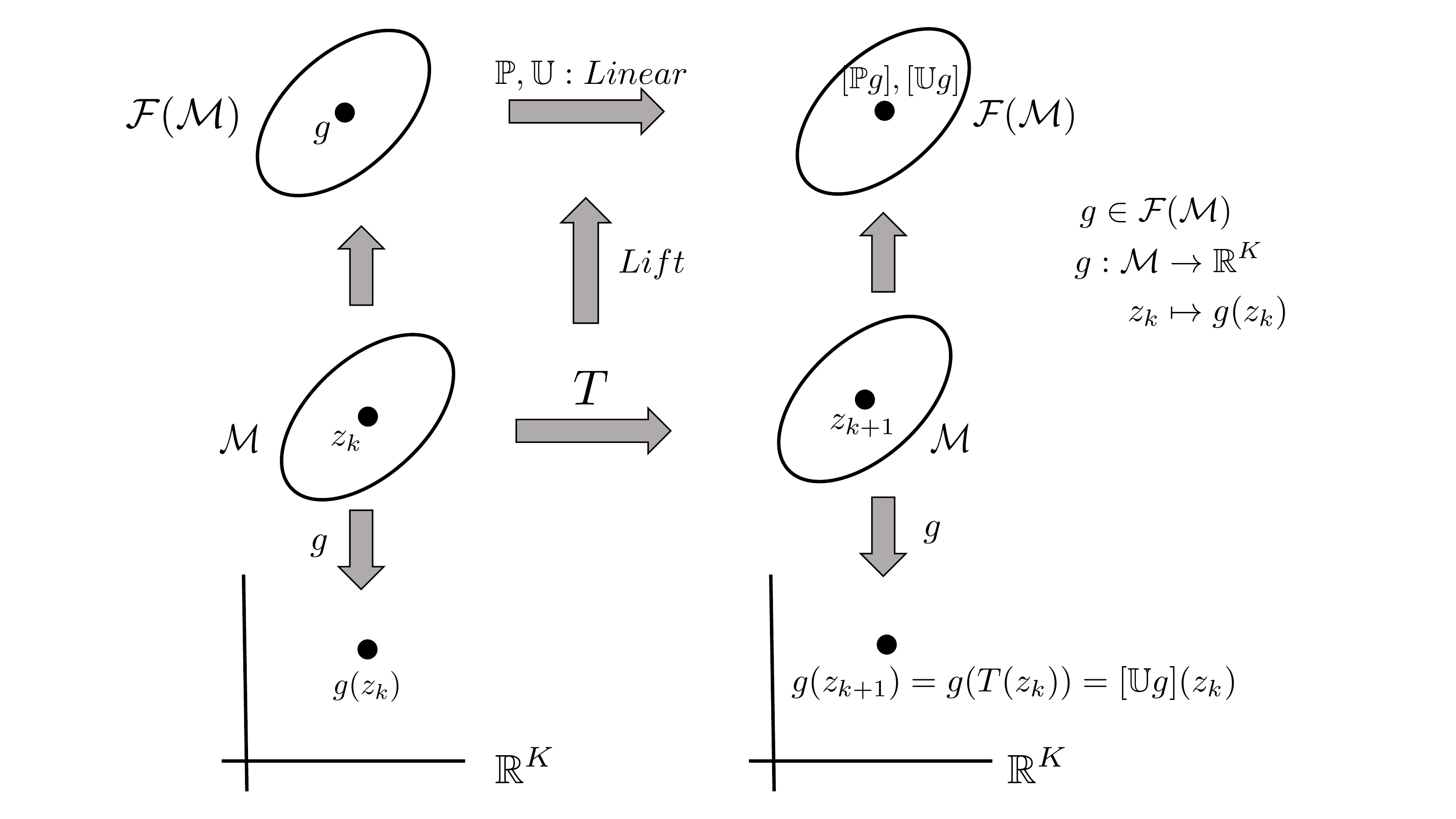}
\caption{Schematic of the P-F and Koopman operators.}\label{koopman_diagram}
\end{figure}

Both the Perron-Frobenius and the Koopman operators are linear operators, even if the underlying system is nonlinear. But while analysis is made tractable by linearity, the trade-off is that these operators are typically infinite dimensional. In particular, the P-F operator and Koopman operator often will lift a dynamical system from a finite-dimensional space to generate an infinite dimensional linear system in infinite dimensions (see Fig. \ref{koopman_diagram}).

\section{Online Koopman Learning of Dynamical Systems}\label{section_online_Koopman}
As mentioned, the Koopman operator is a linear operator, even if the underlying dynamics is nonlinear. The linearity of the operator facilitates the use of linear control techniques for the analysis and control of nonlinear systems. Moreover, Koopman operator analysis is tailor-made for data-driven analysis of dynamical systems. As such, there are several algorithms for computation of the finite-dimensional approximations of the Koopman operators from time-series data with Dynamic Mode Decomposition (DMD) \cite{DMD_schmitt} and Extended Dynamic Mode Decomposition (EDMD) \cite{EDMD_williams} being the most popular ones. 

\subsection{Extended Dynamic Mode Decomposition}
In this subsection, we briefly describe the EDMD algorithm for the finite-dimensional approximation of the Koopman operator.

Consider snap-shots of time-series data as shown below
\begin{eqnarray}
X_p = [x_1,x_2,\ldots,x_M],& X_f = [y_1,y_2,\ldots,y_M] \label{data}
\end{eqnarray}
obtained from simulating a dynamical system $x\mapsto T(x)$, or from an experiment, where $x_i\in X$ and $y_i\in X$, $X\subset \mathbb{R}^N$. The two pair of data sets are assumed to be two consecutive snapshots i.e., $y_i=T(x_i)$. Let $\mathcal{D}=
\{\psi_1,\psi_2,\ldots,\psi_K\}$ be the set of observables, where $\psi : X \to \mathbb{C}$. Let ${\cal G}_{\cal D}$ denote the span of ${\cal D}$ such that ${\cal G}_{\cal D}\subset {\cal G}$, where ${\cal G} = L_2(X)$. Define vector valued function $\mathbf{\Psi}:X\to \mathbb{R}^{K}$
\begin{equation}
\mathbf{\Psi}(\boldsymbol{x}):=\begin{bmatrix}\psi_1(x) & \psi_2(x) & \cdots & \psi_K(x)\end{bmatrix}^\top.
\end{equation}

Here $\mathbf{\Psi}$ is the mapping from physical space to feature space. Any function $\phi,\hat{\phi}\in \mathcal{G}_{\cal D}$ can be written as
\begin{eqnarray}
\phi = \sum_{k=1}^K a_k\psi_k=\boldsymbol{\Psi^T a},\quad \hat{\phi} = \sum_{k=1}^K \hat{a}_k\psi_k=\boldsymbol{\Psi^T \hat{a}}
\end{eqnarray}
for some set of coefficients $\boldsymbol{a},\boldsymbol{\hat{a}}\in \mathbb{R}^K$. Let \[ \hat{\phi}(x)=[\mathbb{U}\phi](x)+r,\]
where $r$ is a residual that appears because $\mathcal{G}_{\cal D}$ is not necessarily invariant to the action of the Koopman operator. The finite dimensional approximate Koopman operator $\bf K$ minimizes this residual $r$ and the matrix $\bf K$ is obtained as a solution of the following least square problem: 
\begin{equation}\label{edmd_op}
\min\limits_{\bf K}\parallel{\bf K} {Y_p}-{Y_f}\parallel_F
\end{equation}
where
\begin{eqnarray}\label{edmd1}
\begin{aligned}
& {Y_p}={\bf \Psi}(X_p) = [{\bf \Psi}(x_1), {\bf \Psi}(x_2), \cdots , {\bf \Psi}(x_M)]\\
& {Y_f}={\bf \Psi}(X_f) = [{\bf \Psi}(y_1), {\bf \Psi}(y_2), \cdots , {\bf \Psi}(y_M)],
\end{aligned}
\end{eqnarray}
with ${\bf K}\in\mathbb{R}^{K\times K}$. The optimization problem (\ref{edmd_op}) can be solved explicitly to obtain following solution for the matrix $\bf K$
\begin{eqnarray}
{\bf K}={Y_f}{Y_p}^\dagger \label{EDMD_formula}
\end{eqnarray}
where ${Y_p}^{\dagger}$ is the pseudo-inverse of matrix $Y_p$.
DMD is a special case of EDMD algorithm with ${\bf \Psi}(x) = x$.

\subsection{Recursive Koopman Learning}
The Koopman operator is generally computed from the formula (\ref{EDMD_formula}), where one uses the entire data-set for the computation. Hence, when a new data point is acquired, the Koopman operator is recomputed using the enlarged data-set. As can be seen from (\ref{EDMD_formula}), computation of the Koopman operator involves the inversion of a matrix and hence, for large dimensional systems, the Koopman operator computation becomes computationally expensive. This issue becomes even more detrimental for real-time identification of dynamical systems, where the Koopman operator needs to be updated with the acquisition of each new data point. This is because with each new data point, the Koopman operator needs to be recomputed and thus, at each time step one has to perform a matrix inversion for the Koopman operator computation. This warrants a recursive algorithm for the computation of the Koopman operator. 

Let
\begin{eqnarray}
^MX_p = [x_1,x_2,\ldots,x_M],& ^MX_f = [y_1,y_2,\ldots,y_M] \label{data_m}
\end{eqnarray}
be $M$ data points obtained from simulation of a dynamical system $x\mapsto T(x)$ or from an experiment, where $y_i=T(x_i)$. Let
\begin{eqnarray}\label{data_m_lifted}
\begin{aligned}
& ^M{Y_p}={\bf \Psi}(X_p) = [{\bf \Psi}(x_1), {\bf \Psi}(x_2), \cdots , {\bf \Psi}(x_M)]\\
& ^M{Y_f}={\bf \Psi}(X_f) = [{\bf \Psi}(y_1), {\bf \Psi}(y_2), \cdots , {\bf \Psi}(y_M)],
\end{aligned}
\end{eqnarray}
be the data points in the lifted space $(\mathbb{R}^K)$, where the points $x_i$ and $y_i$ are mapped by the dictionary functions ${\bf \Psi}$. Let
\begin{eqnarray}\label{Koopman_m_step}
{\bf K}_M = ^M{Y_f}^M{Y_p}^\dagger
\end{eqnarray}
be the Koopman operator obtained by using the $M$ data points. Now, a new data point $(x_{M+1},y_{M+1})$ is aquired. The problem is to update the Koopman operator $\K_M$ to ${\bf K}_{M+1}$, without explicitly computing the inverse $(^{M+1}{Y_p})^\dagger$.

Note that (\ref{Koopman_m_step}) can be rewritten as
\begin{eqnarray}
\K_M \phi_M = z_M
\end{eqnarray}
where 
\begin{eqnarray}
\begin{aligned}
& \phi_M = ^MY_p(^MY_p)^\top=\sum_{i=1}^MY_p^i(Y_p^i)^\top\\
& z_M = ^MY_f(^MY_p)^\top=\sum_{i=1}^MY_f^i(Y_p^i)^\top
\end{aligned}
\end{eqnarray}
and $Y_p^i$ and $Y_f^i$ are $i^{th}$ columns of $^MY_p$ and $^MY_f$ respectively.
Moreover, the updated Koopman operator $\K_{m+1}$ satisfies
\begin{eqnarray}\label{Koopman_updated}
\K_{M+1}\phi_{M+1} = z_{M+1}
\end{eqnarray}

The idea now is to express $\phi_{M+1}$ and $z_{M+1}$ in terms of $\phi_{M}$ and $z_{M}$. In doing so, we obtain  
\[\phi_{M+1} = \phi_M + Y_p^{M+1}(Y_p^{M+1})^\top.\]
%
%
Hence, using the Matrix Inversion Lemma, we get
\begin{eqnarray}\label{phi_iterate}
\phi_{M+1}^{-1} = \phi_M^{-1} - \frac{\phi_M^{-1}Y_p^{M+1}(Y_p^{M+1})^\top\phi_M^{-1}}{1 + (Y_p^{M+1})^\top \phi_M^{-1} Y_p^{M+1}}.
\end{eqnarray}
Moreover,
\begin{eqnarray}\label{zm_iterate}
z_{M+1} = \sum_{i=1}^{M+1}Y_f^i(Y_p^i)^\top=z_M + Y_f^{M+1}(Y_p^{M+1})^\top.
\end{eqnarray}
Hence, from (\ref{Koopman_updated}), the updated Koopman operator is computed as follows. 
\begin{eqnarray}\label{Koopman_new}\nonumber
\K_{M+1} &=& z_{M+1}\phi_{M+1}^{-1}\\ \nonumber
&=& \left(z_M + Y_f^{M+1}(Y_p^{M+1}\right)^\top)\times\\
&& \left(\phi_M^{-1} - \frac{\phi_M^{-1}Y_p^{M+1}(Y_p^{M+1})^\top\phi_M^{-1}}{1 + (Y_p^{M+1})^\top \phi_M^{-1} Y_p^{M+1}}\right).
\end{eqnarray}

Equation (\ref{Koopman_new}) gives the formula for updating the Koopman operator as new data streams in, without explicitly computing the inverse at every step, thus reducing the computational cost and hence improving efficiency.

\subsection{Initialization of the Algorithm}

Equation (\ref{Koopman_new}) gives the updated Koopman $\K_{M+1}$ operator in terms of quantities computed from the previous time step. Hence, for computing the Koopman operator $\K_1$, one needs to initialize both $\phi_0$ and $z_0$. One potential way out of this situation is to compute the Koopman operator $\K_q$ using the initial $q$ data points $(x_i,y_i)$, $i=1,2,\cdots ,q$, $q<M$ as
\[\K_q = ^qY_f ^qY_p^\dagger\]
and use the corresponding $\phi_q$ and $z_q$ to compute the updated Koopman operators $\K_n$, $n>q$. However, one major issue of this approach is the invertibility of $\phi_q$, as for most practical purposes and applications, one would like $q$ to be small and this will imply that $\phi_q$ won't be of full rank, thus resulting in erroneous computation of the Koopman operator. To resolve this issue and to be more suitable for practical applications, we set 
\[\phi_0 = \delta I_K, \quad z_0 = 0_K,\]
where $\delta >0$, $I_K$ is the $K\times K$ identity matrix and $0_K$ is the $K\times K$ zero matrix.

\begin{remark}
Choosing the initialization parameter $\delta$ can be tricky and usually one should run the algorithm multiple times, with different $\delta$, on a given training data-set and choose the one which has the lowest error on some validation data-set. 
\end{remark}

\begin{algorithm}[htp!]
\caption{Algorithm for online Koopman Operator computation using streaming data.}
\begin{enumerate}
\item{Fix the dictionary functions $\bf \Psi$.}
\item{Initialize $\phi_0 = \delta I_K$ and $ z_0 = 0_K$.}
\item{As a new data point $(x_{M},y_{M})$ streams in, lift the data point to $\mathbb{R}^K$ using the dictionary function $\bf \Psi$.}
\item{Update $z_{M}$ and $\phi_{M}^{-1}$ as
\begin{eqnarray*}
z_{M} &=& z_{M-1} + Y_f^{M}(Y_p^{M})^\top\\
\phi_{M}^{-1} &=& \phi_{M-1}^{-1} - \frac{\phi_{M-1}^{-1}Y_p^{M}(Y_p^{M})^\top\phi_{M-1}^{-1}}{1 + (Y_p^{M})^\top \phi_{M-1}^{-1} Y_p^{M}}.
\end{eqnarray*}
}
\item{Update the Koopman operator $\K_{M-1}$ to $\K_{M}$ as 
\begin{eqnarray*}
\K_{M} = z_{M}\phi_{M}^{-1}.
\end{eqnarray*}
}
\end{enumerate}
\label{algo}
\end{algorithm}

\section{Design of Koopman Predictor}\label{section_predictor}


The linearity of Perron-Frobenius and Koopman operators can be used to design predictors for the underlying system. In this section, we briefly present the predictor design problem for the self-containment of the paper. For details, we refer the readers to \cite{korda_mezic_predictor}. 

Let $\K_M$ be the Koopman operator computed from a streaming data-set $[x_1,\cdots , x_{M+1}]$, using algorithm \ref{algo} and let $\bar x_0$ be the initial condition from where the future trajectory needs to be predicted. Let  
 \[{\bf \Psi}(\bar x_0)=: {\bf z}_0\in \mathbb{R}^K\]
 be the data point in the lifted space. Then the initial condition is propagated using Koopman operator as \[{\bf z}_n={\bf K}_M^n{\bf z}_0.\] The predicted trajectory in the state space is then obtained as 
\[\bar x_n=C {\bf z}_n\]
where matrix $C$ is obtained as the solution of the following least squares problem
\begin{eqnarray}\label{C_pred}
\min_C\sum_{i = 1}^{M+1} \parallel x_i - C \boldsymbol \Psi (x_i)\parallel_2^2
\end{eqnarray}

\section{Iterative Identification of Power Networks}\label{section_simulations}
In this section, the proposed recursive EDMD algorithm is applied to identify the dynamics of an IEEE 9 bus and 39 bus power network. All the simulations were performed in MATLAB\_R2018b on an Apple Macbook Pro with 2.3 GHz Intel Core i5 processor and 8 GB 2133 MHz LPDDR3 RAM. 

\subsection{IEEE 39 Bus System}
\begin{figure}[htp!]
\centering
\includegraphics[scale=.25]{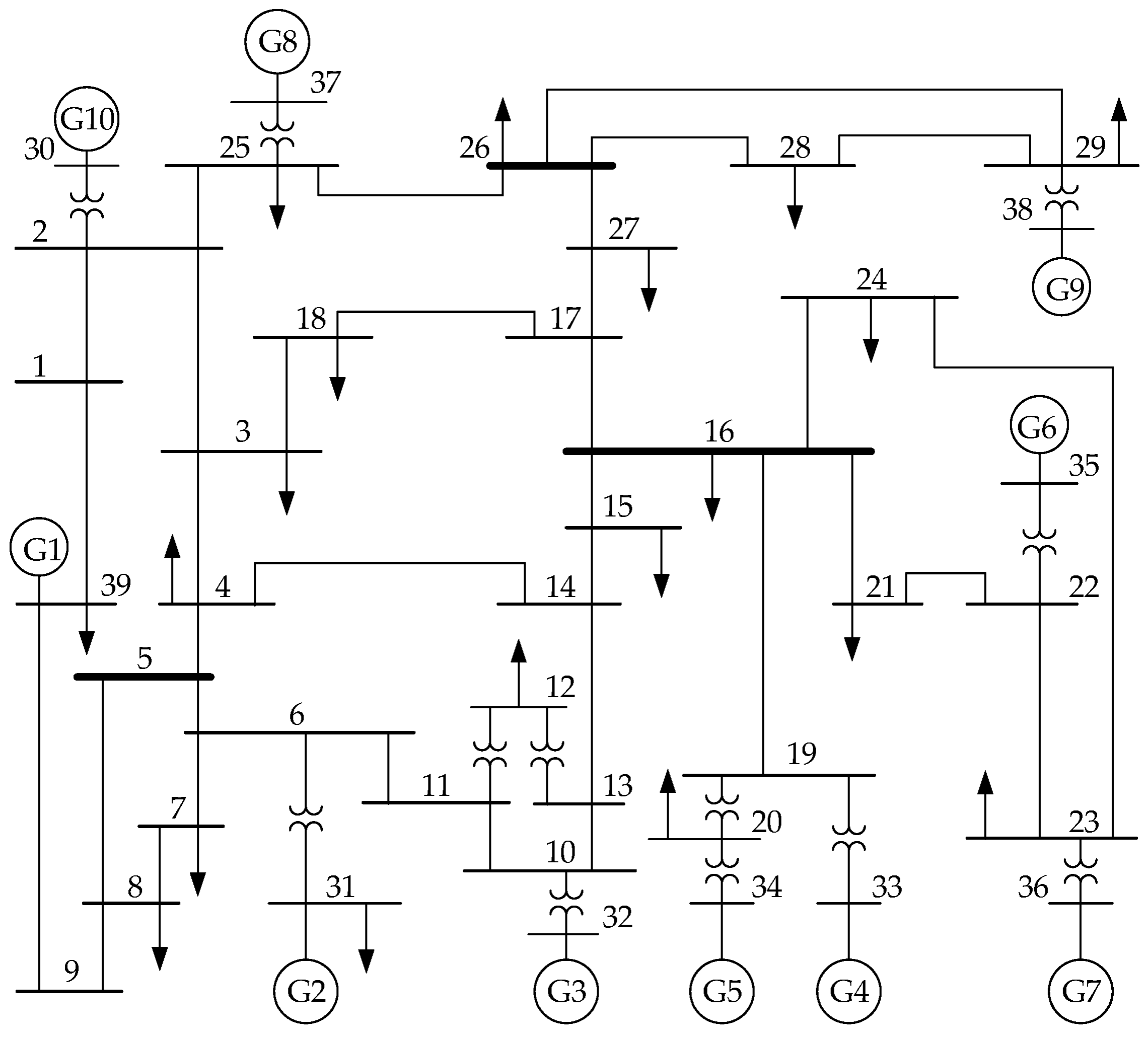}
\caption{One-line representation of IEEE 39 bus system.}\label{39_bus_fig}
\end{figure}

In this subsection we identify the dynamics of IEEE 39 bus network (refer Fig. \ref{39_bus_fig}). 

In this example, we modelled the generator as a $4^{th}$ order system with a $3^{rd}$ order PSS at each generator. The detailed model can be found in \cite{Sauer_pai_book}. We considered data from the linearized model and computed the recursive Koopman operator using linear dictionary functions (DMD algorithm).

\begin{figure}[htp!]
\centering
\subfigure[]{\includegraphics[scale=.2]{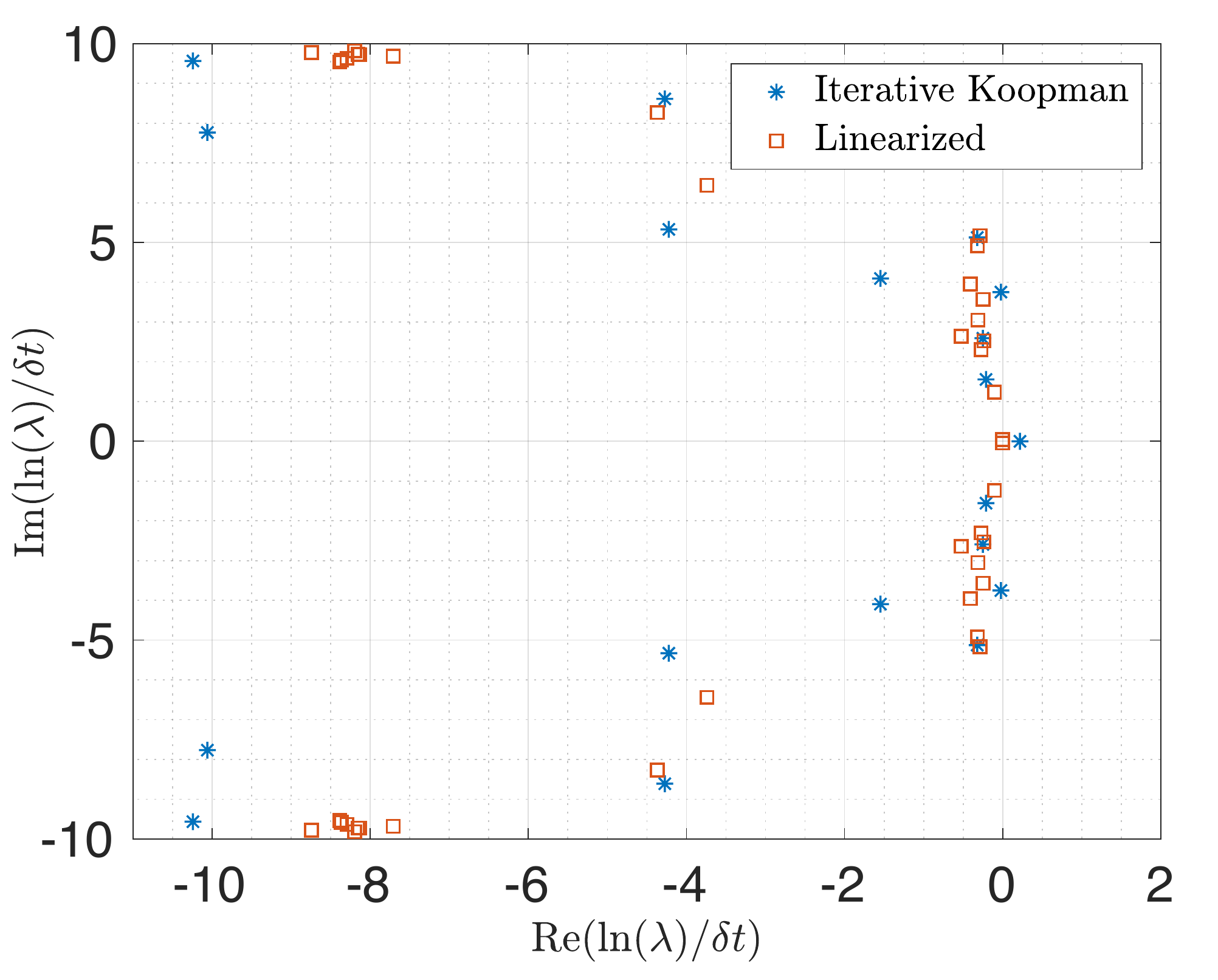}}
\subfigure[]{\includegraphics[scale=.2]{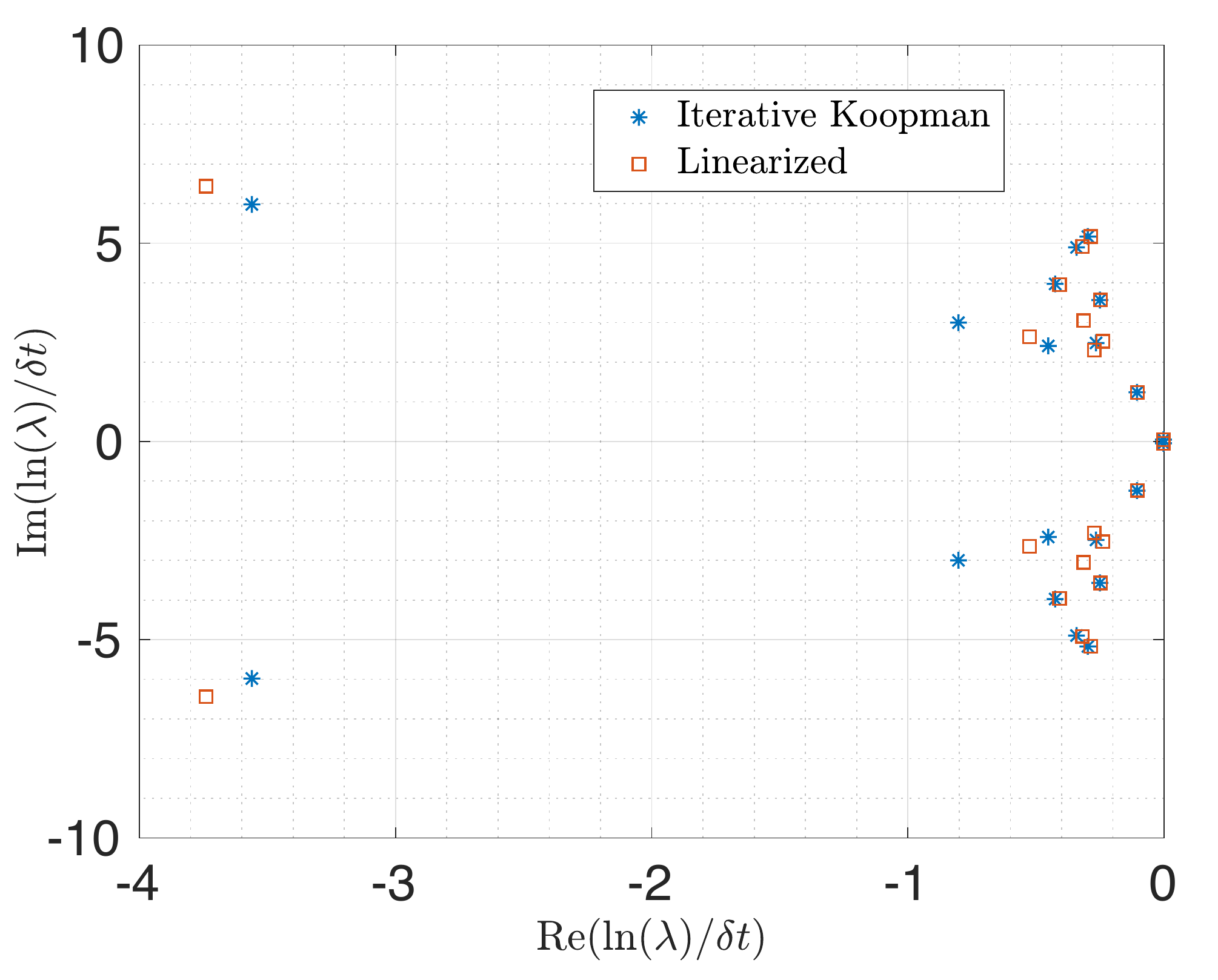}}
\subfigure[]{\includegraphics[scale=.214]{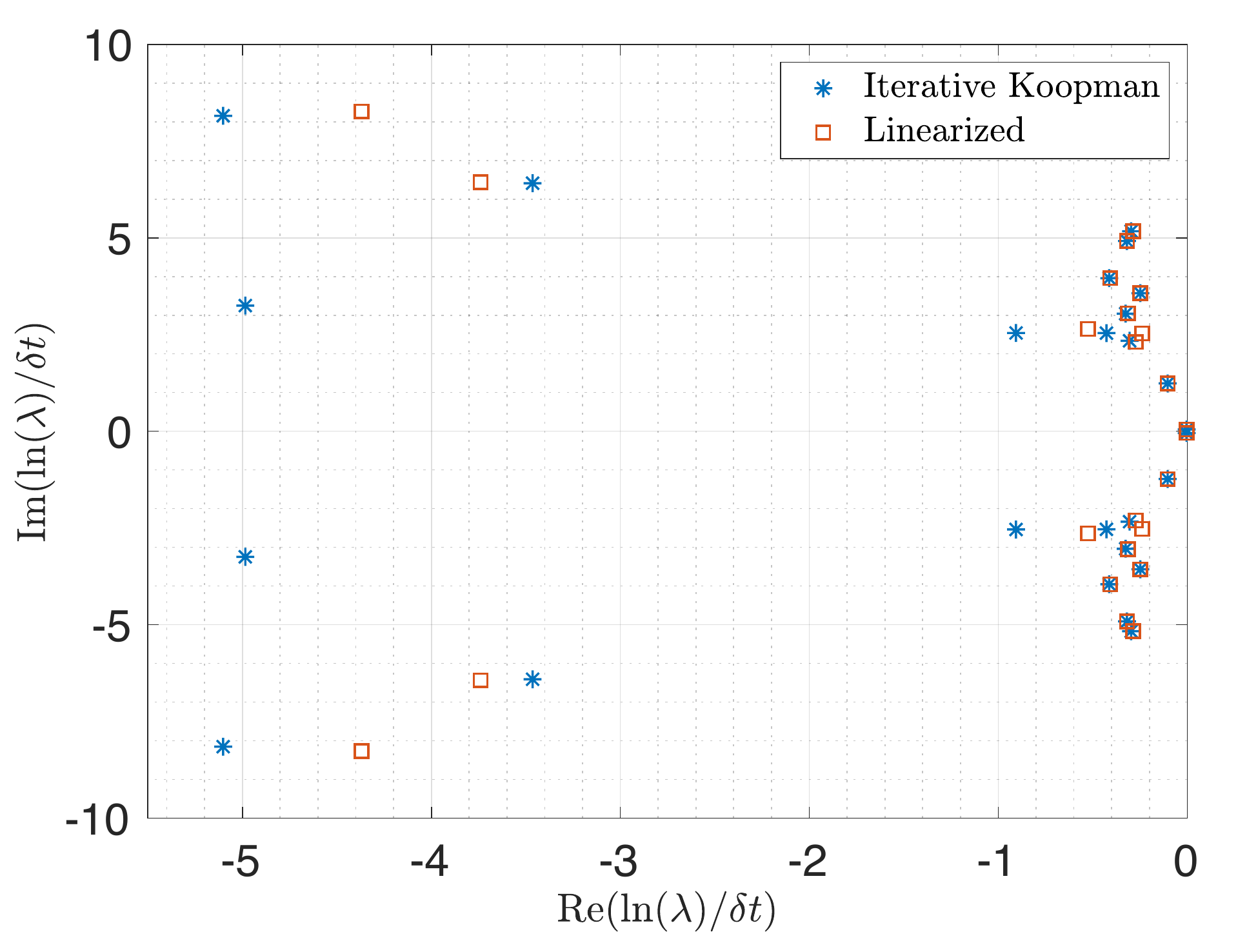}}
\subfigure[]{\includegraphics[scale=.2]{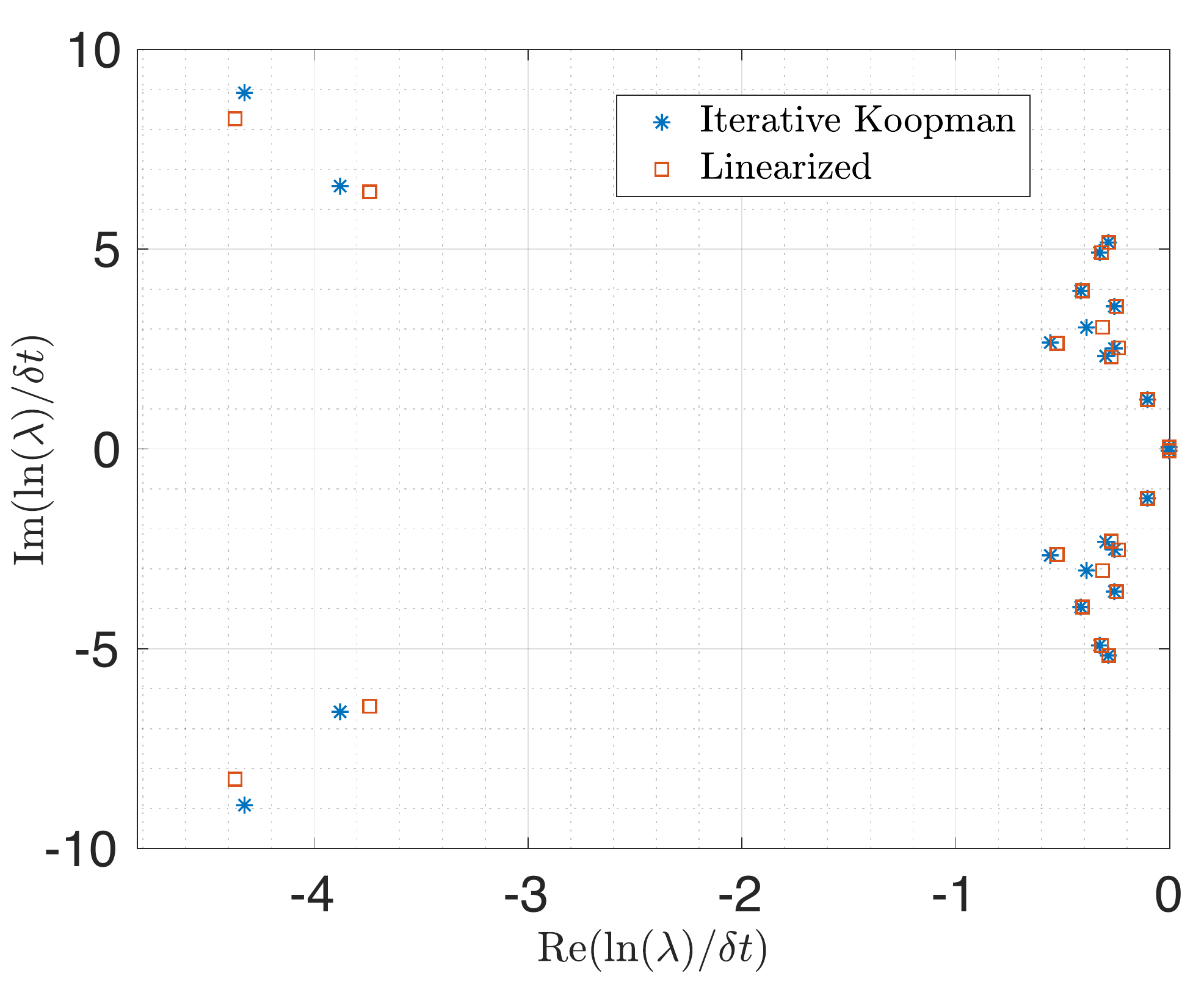}}
\caption{Comparison of dominant Eigenvalues of iterative Koopman operator and dominant eigenvalues of linearized dynamics for (a) $50$ time steps, (b) $100$ time steps, (c) $300$ time steps, (d) $1000$ time steps.}\label{39_bus_eig_fig}
\end{figure}

The evolution of the dominant eigenvalues of the Koopman operator, as the number of data points are increased, is shown in Fig. \ref{39_bus_eig_fig}. It can be seen that as the size of training data set increases, the Koopman eigenvalues approach the eigenvalues of the linearized system and with 1000 training data points, there is almost a perfect match of the dominant eigenvalues between the Koopman operator and the linearized dynamics.

\begin{figure}[htp!]
\centering
\includegraphics[scale = 0.225]{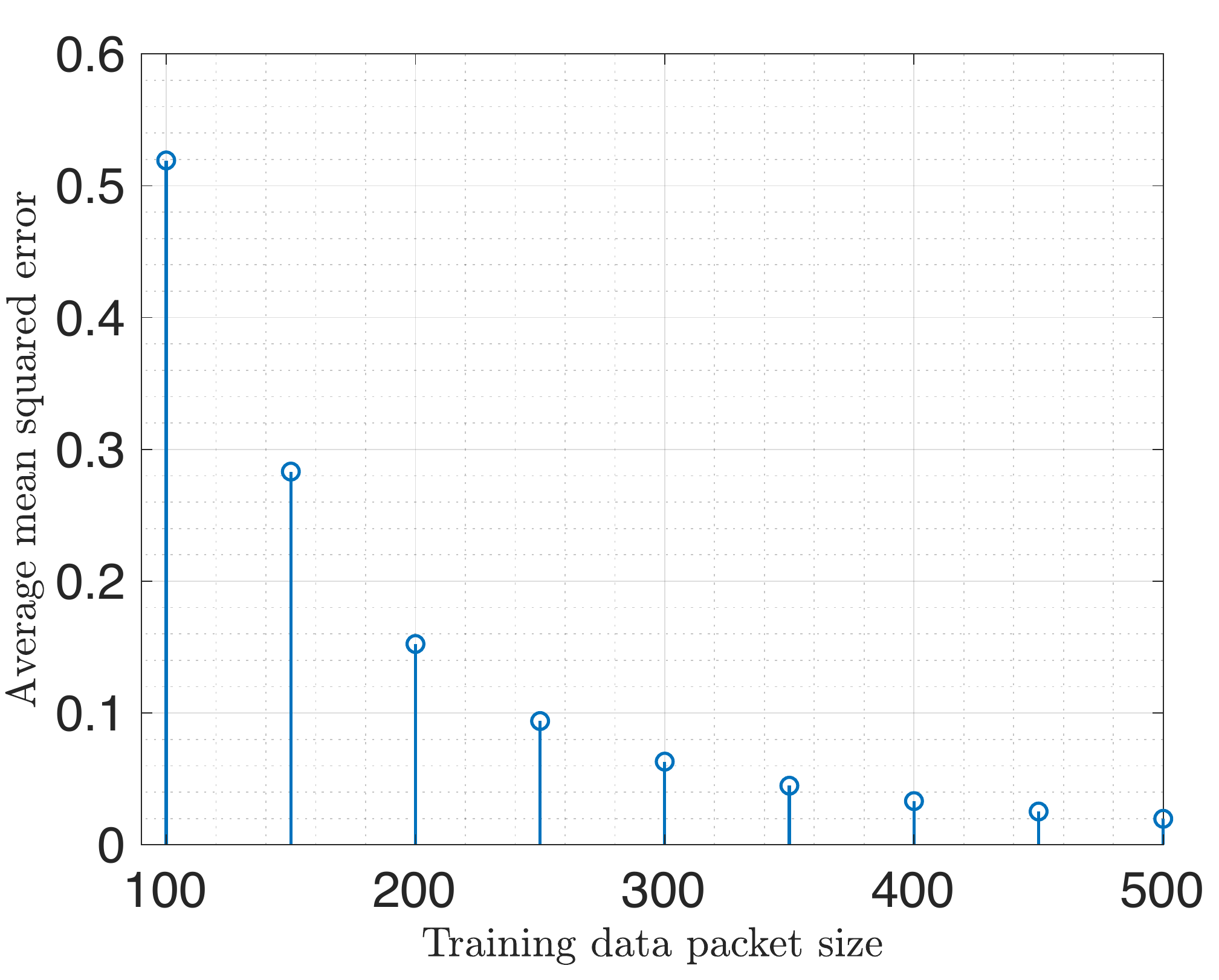}
\caption{Mean square error in prediction of the states with varying training data size.}\label{mse_err_fig}
\end{figure}

Using the different Koopman operators obtained from different sizes of the training data-set, we also predicted the future of the states from time step $600$ to $900$. We used training data up to time step $500$ and since we predict the states from time $600$ to $900$, the learned Koopman is being tested on a data-set which the algorithm has not seen. The corresponding mean square errors in the prediction of the states, as the size of the training data-set varies, is shown in Fig. \ref{mse_err_fig}.

As is expected from intuition, if the Koopman operator is trained using more and more data, the accuracy of prediction increases. Further, as in the 9 bus example, we compared the computation time of our algorithm against the existing DMD framework and the computation times are given in Table \ref{39_bus_comp_time}.

{\small
\begin{table}[htp!]
\centering
\caption{Comparison of computation time for REDMD and EDMD}\label{39_bus_comp_time}
\begin{tabular}{|c|c|c|c|}
\hline
\multirow{3}{*}{\# of Data points} & \multirow{2}{*}{REDMD} & \multirow{2}{*}{EDMD} \\
                      &                        &                       \\
                      & computational time (s)                   & computational time (s)                  \\ \hline
    $50$ & $0.0407$ & $0.0610$ \\
     $100$ & $0.0897$ & $0.1547$\\
     $300$ & $0.2070$  & $0.3723$\\
     $500$ & $0.4105$ & $0.7998$\\
     $1000$ & $0.6358$ & $2.1848$\\
\hline
\end{tabular}
\end{table}
}
As before our algorithm is faster than the existing DMD algorithm. However, it can be seen that the difference in computation time in this case is not as much as in the 9 bus example. This is because, in this example we chose linear dictionary functions, whereas in the 9 bus example we had chosen 150 Gaussian radial basis function as dictionary functions. 

\subsection{IEEE 9 bus System}
\begin{figure}[htp!]
\centering
\includegraphics[scale=1.2]{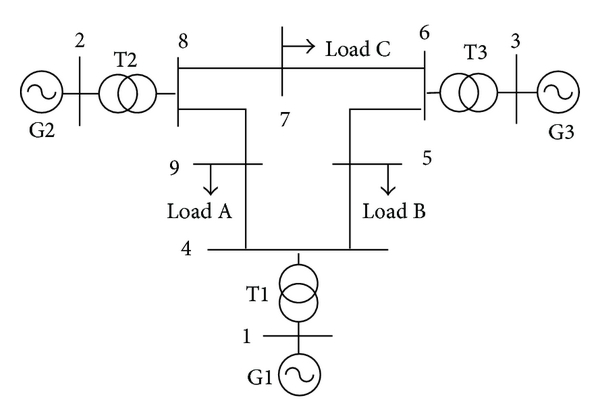}
\caption{One-line representation of IEEE 9 bus system.}\label{9_bus_fig}
\end{figure}

We consider the IEEE 9 bus system as shown in Fig. \ref{9_bus_fig}. 

The network has three generators and a $6^{th}$ order model for each generator is considered, an IEEE Type I exciter for each generator consisting of a $3^{rd}$ order dynamic model and a type II governor model is considered for each generator with first-order dynamics. Hence, the resultant system is a $30$ dimensional nonlinear system. The detailed modeling of the generator can be found in \cite{power_model} and is omitted in this paper due to page constraints. For the analysis, the data is generated using PSAT \cite{milano2005open} in MATLAB by perturbing the system from a steady state. The data is sampled at a frequency of $0.01$ seconds, which is in accordance with PMU measurements. 

\begin{figure}[htp!]
\centering
\subfigure[]{\includegraphics[scale=.187]{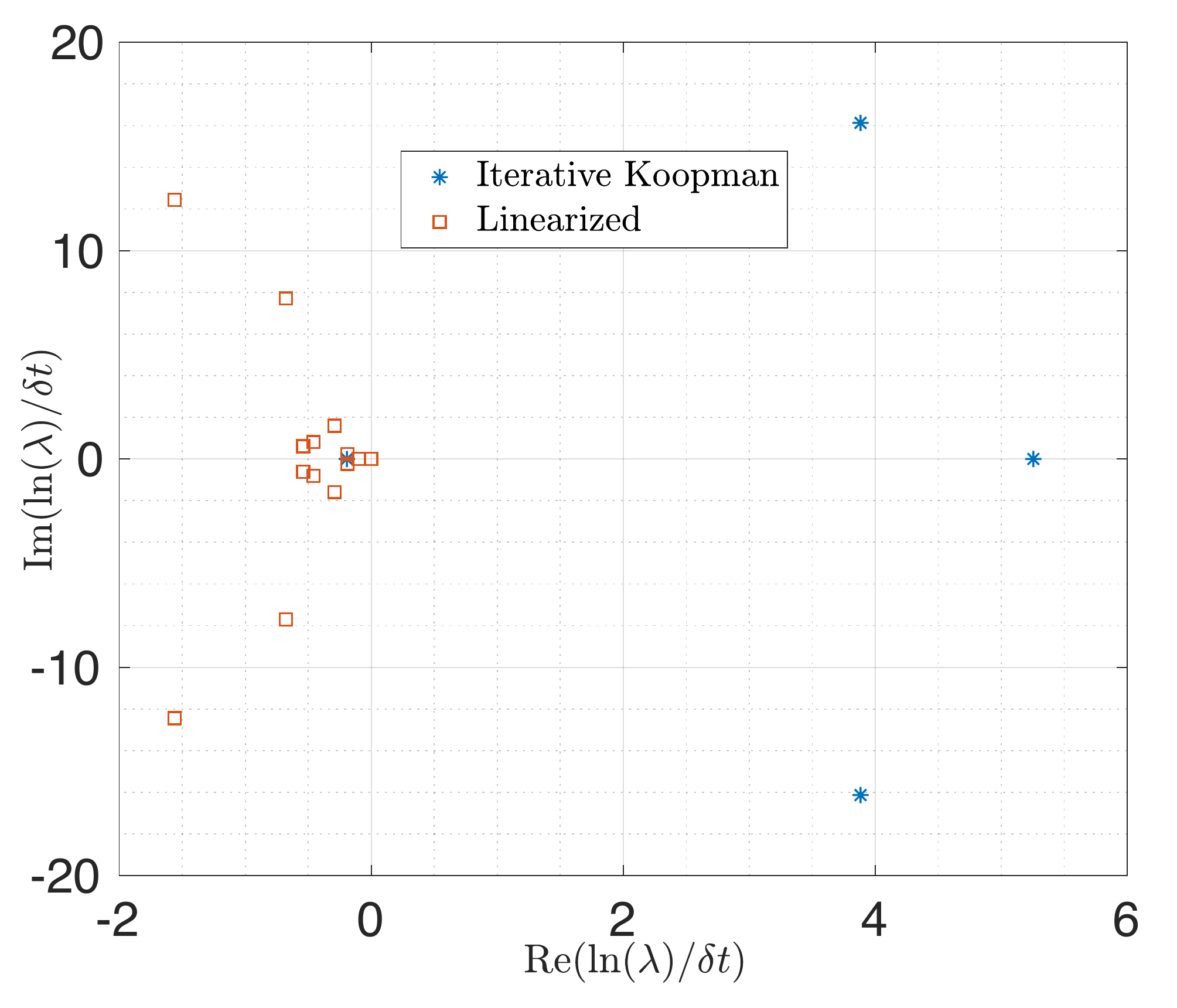}}
\subfigure[]{\includegraphics[scale=.2]{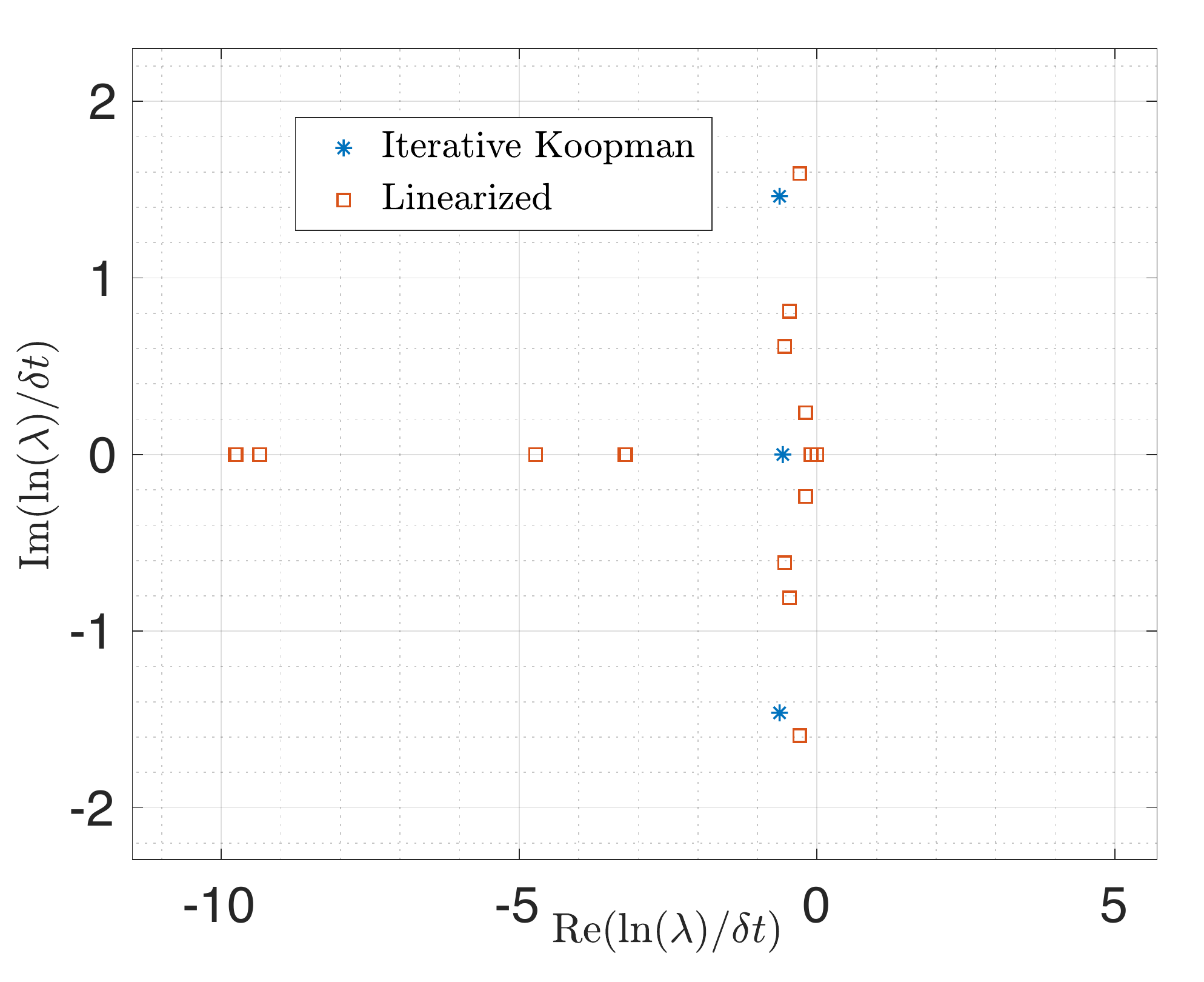}}
\subfigure[]{\includegraphics[scale=.2]{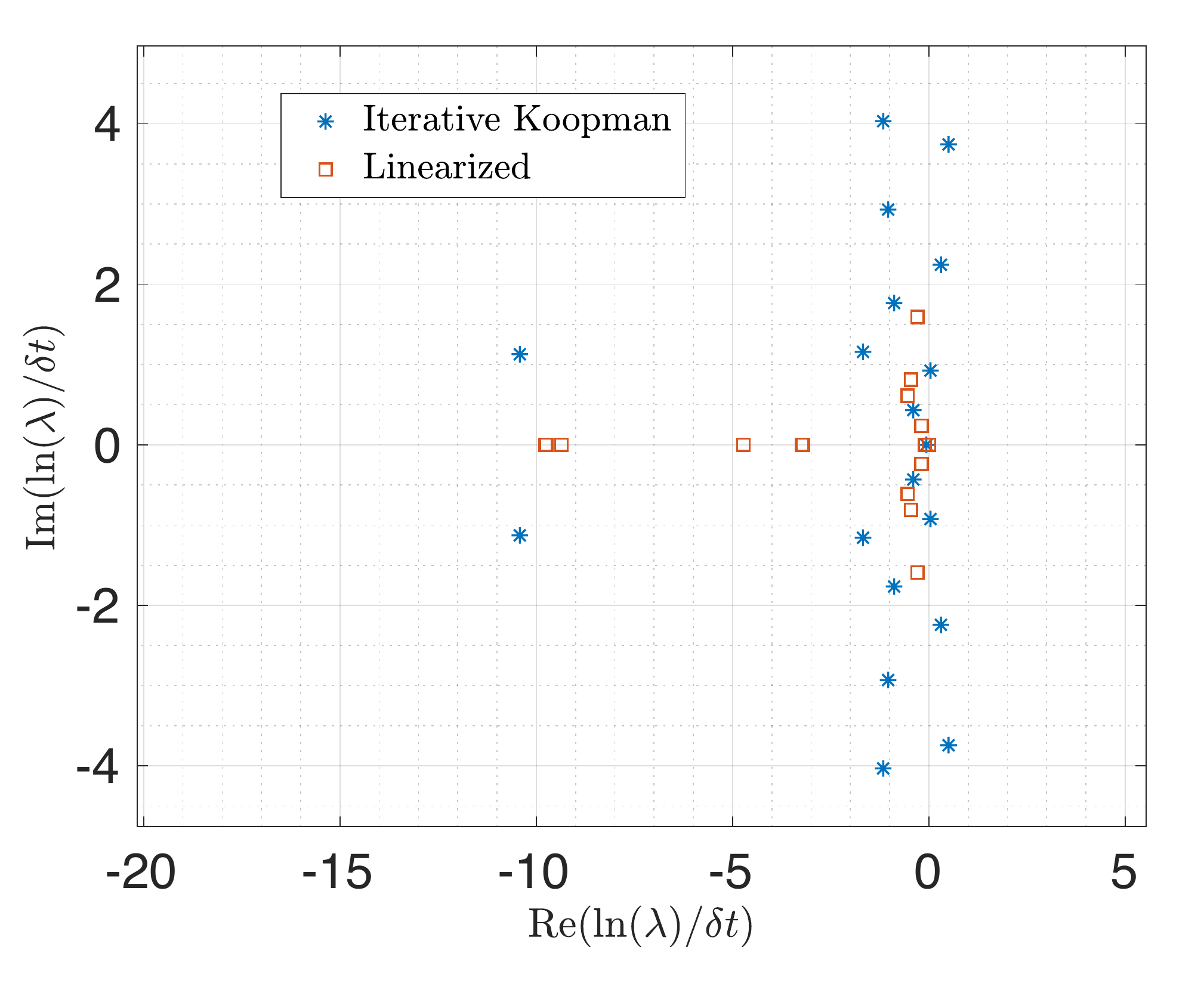}}
\subfigure[]{\includegraphics[scale=.2]{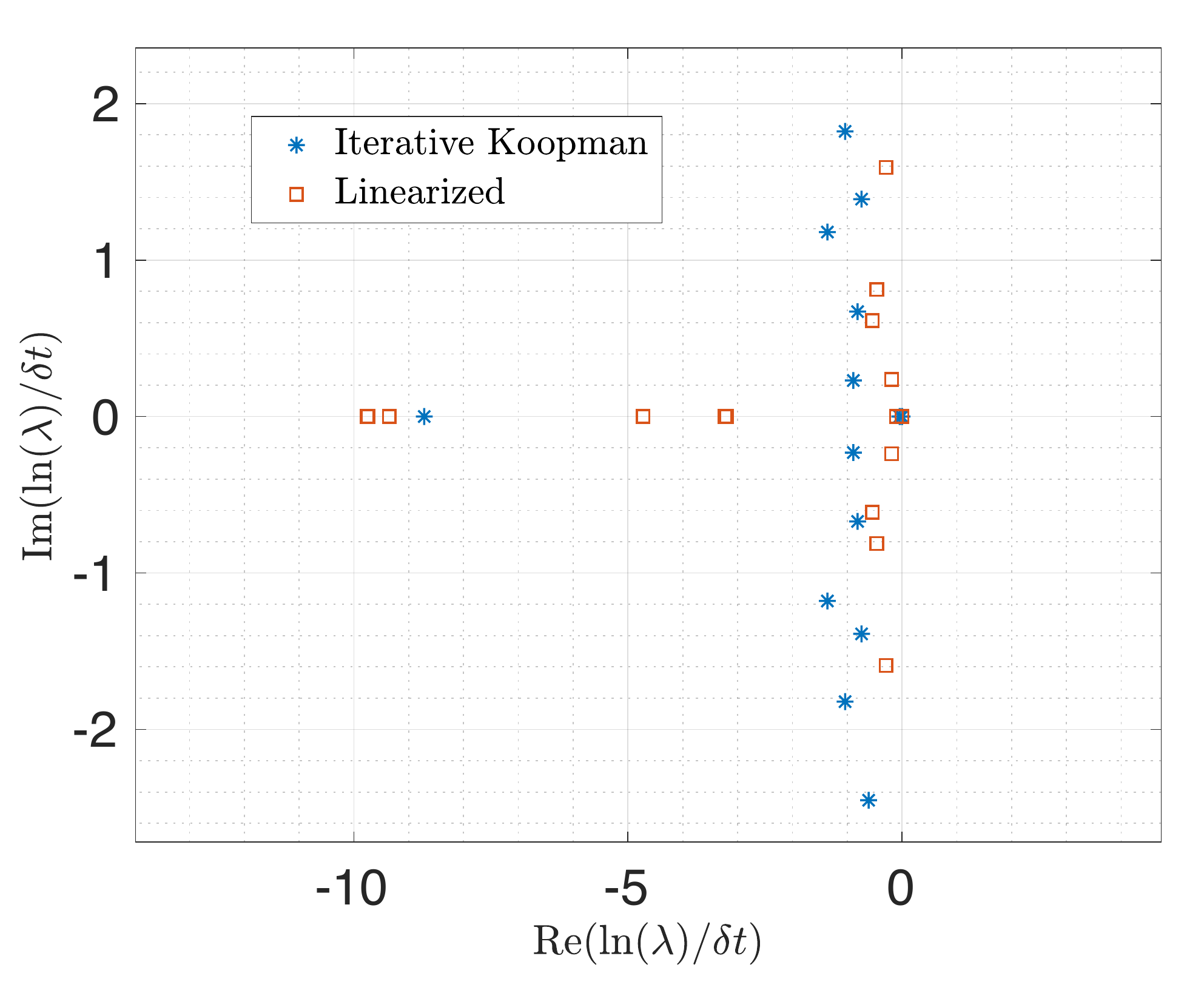}}
\caption{Comparison of dominant Eigenvalues of recursive Koopman operator and dominant eigenvalues of linearized dynamics for (a) 250 time steps, (b) 500 time steps, (c) 1000 time steps, (d) 2000 time steps.}\label{9_bus_eig_fig}
\end{figure}

Unlike the previous test case, in this example we consider nonlinear data and use Recursive EDMD algorithm, with 150 Gaussian radial basis functions as the dictionary functions ${\bf \Psi}$. With this, we computed the Koopman operator iteratively for 2000 time steps and the dominant eigenvalues of the recursive Koopman operator are plotted in Fig. \ref{9_bus_eig_fig}.

It can be seen from Fig. \ref{9_bus_eig_fig}(a) that when the data set is small in size, the Koopman operator has positive eigenvalues. This is a drawback of the Koopman framework and has been reported in the literature \cite{robust_DMD_arxiv,sparse_Koopman_acc}. But as more data points are acquired, the dominant eigenvalues of the Koopman operator approach the dominant eigenvalues of the linearized system (see Fig. \ref{9_bus_eig_fig}(b)-(d)).

However, the main novelty of the proposed approach lies in the fact that our proposed algorithm reduces the computation time by a large extent when new data-points are acquired from sensors in real-time and thus making a real-time update of the Koopman operator, necessary for real-time system identification. In Table \ref{9_bus_comp_time}, we compare the computation times of the Koopman operator using our proposed iterative method with the existing EDMD method. It can be seen that the computation time of the desired operator and its eigenspectrum reduces drastically. Moreover, the reduction in computation time is nonlinear and with larger systems, the difference in computation time will be even greater. 



{\small
\begin{table}[htp!]
\centering
\caption{Comparison of computation time for REDMD and EDMD}\label{9_bus_comp_time}
\begin{tabular}{|c|c|c|}
\hline
\multirow{3}{*}{\# of Data points} & \multirow{2}{*}{REDMD} & \multirow{2}{*}{EDMD} \\
                      &                        &                       \\
                      & computational time (s)                   & computational time (s)                  \\ \hline
$250$                      &    $1.832$ & $2.287$                  \\ 
$500$                      &    $3.984$ & $7.127$              \\ 
$1000$                     &    $5.596$ & $12.115$             \\ 
$1500$                     &    $9.483$ & $19.907$             \\ 
$2000$                     &    $12.886$ & $30.169$            \\ \hline
\end{tabular}
\end{table}
}

\section{Conclusions}\label{section_conclusion}

In this paper, a novel algorithm for online learning of power system dynamics from real-time streaming data is proposed. In particular, we use operator theoretic ideas, namely Koopman operator to approximate the underlying dynamics of a power network using data. The proposed approach is computationally efficient when compared to existing popular approaches such as EDMD and this facilitates the real-time learning of power system dynamics. The efficacy of the proposed approach is demonstrated on two different power networks, namely IEEE 9 bus system with nonlinear dynamics and IEEE 39 bus system with linearized dynamics. Real time learning of power system dynamics enables better monitoring capabilities and future efforts involve anomaly detection and classification from the linearized higher dimensional power system dynamics.


\bibliographystyle{IEEEtran}
\bibliography{subhrajit_online_DMD_PES}

\begin{thebibliography}{10}
\providecommand{\url}[1]{#1}
\csname url@samestyle\endcsname
\providecommand{\newblock}{\relax}
\providecommand{\bibinfo}[2]{#2}
\providecommand{\BIBentrySTDinterwordspacing}{\spaceskip=0pt\relax}
\providecommand{\BIBentryALTinterwordstretchfactor}{4}
\providecommand{\BIBentryALTinterwordspacing}{\spaceskip=\fontdimen2\font plus
\BIBentryALTinterwordstretchfactor\fontdimen3\font minus
  \fontdimen4\font\relax}
\providecommand{\BIBforeignlanguage}[2]{{%
\expandafter\ifx\csname l@#1\endcsname\relax
\typeout{** WARNING: IEEEtran.bst: No hyphenation pattern has been}%
\typeout{** loaded for the language `#1'. Using the pattern for}%
\typeout{** the default language instead.}%
\else
\language=\csname l@#1\endcsname
\fi
#2}}
\providecommand{\BIBdecl}{\relax}
\BIBdecl

\bibitem{de2010synchronized}
J.~De~La~Ree, V.~Centeno, J.~S. Thorp, and A.~G. Phadke, ``Synchronized phasor
  measurement applications in power systems,'' \emph{IEEE Transactions on smart
  grid}, vol.~1, no.~1, pp. 20--27, 2010.

\bibitem{pushpak2016control}
S.~Pushpak and U.~Vaidya, ``Control of inter-area oscillation with noise
  corrupted wide area measurement,'' in \emph{2016 American Control Conference
  (ACC)}.\hskip 1em plus 0.5em minus 0.4em\relax IEEE, 2016, pp. 7498--7503.

\bibitem{Dellnitz_Junge}
M.~Dellnitz and O.~Junge, ``On the approximation of complicated dynamical
  behavior,'' \emph{SIAM Journal on Numerical Analysis}, vol.~36, pp. 491--515,
  1999.

\bibitem{Mezic2000}
I.~Mezic and A.~Banaszuk, ``Comparison of systems with complex behavior:
  spectral methods,'' in \emph{Proceedings of the 39th IEEE Conference on
  Decision and Control (Cat. No.00CH37187)}, vol.~2, 2000, pp. 1224--1231
  vol.2.

\bibitem{froyland_extracting}
G.~Froyland, ``Extracting dynamical behaviour via {Markov} models,'' in
  \emph{Nonlinear Dynamics and Statistics: Proceedings, Newton Institute,
  Cambridge, 1998}, A.~Mees, Ed.\hskip 1em plus 0.5em minus 0.4em\relax
  Birkhauser, 2001, pp. 283--324.

\bibitem{Junge_Osinga}
O.~Junge and H.~Osinga, ``A set oriented approach to global optimal control,''
  \emph{ESAIM: Control, Optimisation and Calculus of Variations}, vol.~10,
  no.~2, pp. 259--270, 2004.

\bibitem{Mezic_comparison}
I.~Mezi\'{c} and A.~Banaszuk, ``Comparison of systems with complex behavior,''
  \emph{Physica D}, vol. 197, pp. 101--133, 2004.

\bibitem{Dellnitztransport}
M.~Dellnitz, O.~Junge, W.~S. Koon, F.~Lekien, M.~Lo, J.~E. Marsden, K.~Padberg,
  R.~Preis, S.~D. Ross, and B.~Thiere, ``Transport in dynamical astronomy and
  multibody problems,'' \emph{International Journal of Bifurcation and Chaos},
  vol.~15, pp. 699--727, 2005.

\bibitem{mezic2005spectral}
I.~Mezi{\'c}, ``Spectral properties of dynamical systems, model reduction and
  decompositions,'' \emph{Nonlinear Dynamics}, vol.~41, no. 1-3, pp. 309--325,
  2005.

\bibitem{Mehta_comparsion_cdc}
P.~G. Mehta and U.~Vaidya, ``On stochastic analysis approaches for comparing
  dynamical systems,'' in \emph{Proceeding of IEEE Conference on Decision and
  Control}, Spain, 2005, pp. 8082--8087.

\bibitem{Vaidya_TAC}
U.~Vaidya and P.~G. Mehta, ``Lyapunov measure for almost everywhere
  stability,'' \emph{IEEE Transactions on Automatic Control}, vol.~53, no.~1,
  pp. 307--323, 2008.

\bibitem{raghunathan2014optimal}
A.~Raghunathan and U.~Vaidya, ``Optimal stabilization using lyapunov
  measures,'' \emph{IEEE Transactions on Automatic Control}, vol.~59, no.~5,
  pp. 1316--1321, 2014.

\bibitem{susuki2011nonlinear}
Y.~Susuki and I.~Mezic, ``Nonlinear koopman modes and coherency identification
  of coupled swing dynamics,'' \emph{IEEE Transactions on Power Systems},
  vol.~26, no.~4, pp. 1894--1904, 2011.

\bibitem{mezic_koopmanism}
M.~Budisic, R.~Mohr, and I.~Mezic, ``Applied koopmanism,'' \emph{Chaos},
  vol.~22, pp. 047\,510--32, 2012.

\bibitem{mezic_koopman_stability}
A.~Mauroy and I.~Mezic, ``A spectral operator-theoretic framework for global
  stability,'' in \emph{Proc. of IEEE Conference of Decision and Control},
  Florence, Italy, 2013.

\bibitem{yeung2015global}
E.~Yeung, J.~Kim, J.~Gon{\c{c}}alves, and R.~M. Murray, ``Global network
  identification from reconstructed dynamical structure subnetworks:
  Applications to biochemical reaction networks,'' in \emph{Decision and
  Control (CDC), 2015 IEEE 54th Annual Conference on}.\hskip 1em plus 0.5em
  minus 0.4em\relax IEEE, 2015, pp. 881--888.

\bibitem{yeung2018koopman}
E.~Yeung, Z.~Liu, and N.~O. Hodas, ``A koopman operator approach for computing
  and balancing gramians for discrete time nonlinear systems,'' in \emph{2018
  Annual American Control Conference (ACC)}.\hskip 1em plus 0.5em minus
  0.4em\relax IEEE, 2018, pp. 337--344.

\bibitem{yeung2017learning}
E.~Yeung, S.~Kundu, and N.~Hodas, ``Learning deep neural network
  representations for koopman operators of nonlinear dynamical systems,''
  \emph{arXiv preprint arXiv:1708.06850}, 2017.

\bibitem{sparse_Koopman_acc}
S.~Sinha, U.~Vaidya, and E.~Yeung, ``On computation of koopman operator from
  sparse data,'' in \emph{2019 American Control Conference (ACC)}.\hskip 1em
  plus 0.5em minus 0.4em\relax IEEE, 2019, pp. 5519--5524.

\bibitem{johnson2018class}
C.~A. Johnson and E.~Yeung, ``A class of logistic functions for approximating
  state-inclusive koopman operators,'' in \emph{2018 Annual American Control
  Conference (ACC)}.\hskip 1em plus 0.5em minus 0.4em\relax IEEE, 2018, pp.
  4803--4810.

\bibitem{sinha_online_arxiv}
S.~Sinha, S.~P. Nandanoori, and E.~Yeung, ``Online learning of dynamical
  systems: An operator theoretic approach,'' \emph{arXiv preprint
  arXiv:1909.12520}, 2019.

\bibitem{sai_phase_space_arxiv}
S.~P. Nandanoori, S.~Sinha, and E.~Yeung, ``Data-driven operator theoretic
  methods for global phase space learning,'' \emph{arXiv preprint
  arXiv:1910.03011}, 2019.

\bibitem{DMD_schmitt}
P.~J. Schmid, ``Dynamic mode decomposition of numerical and experimental
  data,'' \emph{Journal of Fluid Mechanics}, vol. 656, pp. 5--28, 2010.

\bibitem{EDMD_williams}
M.~O. Williams, I.~G. Kevrekidis, and C.~W. Rowley, ``A data--driven
  approximation of the koopman operator: Extending dynamic mode
  decomposition,'' \emph{Journal of Nonlinear Science}, vol.~25, no.~6, pp.
  1307--1346, 2015.

\bibitem{Lasota}
A.~Lasota and M.~C. Mackey, \emph{Chaos, Fractals, and Noise: Stochastic
  Aspects of Dynamics}.\hskip 1em plus 0.5em minus 0.4em\relax New York:
  Springer-Verlag, 1994.

\bibitem{korda_mezic_predictor}
M.~Korda and I.~Mezi{\'c}, ``Linear predictors for nonlinear dynamical systems:
  Koopman operator meets model predictive control,'' \emph{arXiv preprint
  arXiv:1611.03537}, 2016.

\bibitem{Sauer_pai_book}
P.~W. Sauer and M.~Pai, ``Power system dynamics and stability,'' \emph{Urbana},
  vol.~51, p. 61801, 1997.

\bibitem{power_model}
S.~K. Khaitan and J.~D. McCalley, ``High performance computing for power system
  dynamic simulation,'' in \emph{High performance computing in power and energy
  systems}.\hskip 1em plus 0.5em minus 0.4em\relax Springer, 2013, pp. 43--69.

\bibitem{milano2005open}
F.~Milano, ``An open source power system analysis toolbox,'' \emph{IEEE
  Transactions on Power systems}, vol.~20, no.~3, pp. 1199--1206, 2005.

\bibitem{robust_DMD_arxiv}
S.~Sinha, H.~Bowen, and U.~Vaidya, ``On robust computation of koopman operator
  and prediction in random dynamical systems,'' \emph{arXiv preprint
  arXiv:1803.08562}, 2018.

\end{thebibliography}

\end{document}